# Influence of Temperature and Frequency on Electric Field Reduction Method via a Nonlinear Field Dependent Conductivity Layer Combined with Protruding Substrate for Power Electronics Modules

Maryam Mesgarpour Tousi and Mona Ghassemi
ECE Department, Virginia Tech, Blacksburg, VA, USA

*Abstract*—As shown in our previous studies, geometrical field grading techniques such as stacked and protruding substrate designs cannot well mitigate high electric stress issue within power electronics modules. However, it was shown that a combination of protruding substrate design and applying a nonlinear field-dependent conductivity layer could address the issue. Electric field (*E*) simulations were carried out according to IEC 61287-1 for the partial discharge test measurement step, where a 50/60 Hz AC voltage was applied. However, dielectrics, including ceramic substrate and silicone gel, in power devices undergo high temperatures up to a few hundred degrees and frequencies up to 1 MHz. Thus, E values obtained with electrical parameters of the mentioned dielectrics for room temperature and under 50/60 Hz may not be valid for high temperatures and frequencies mentioned above. In this paper, we address this technical gap through developing a finite element method (FEM) *E* calculation model developed in COMSOL Multiphysics where *E* calculations are carried out for different temperatures up to 250ºC and frequencies up to 1 MHz. Using the model, the influence of temperature and frequency on our proposed electric field mitigation technique mentioned above is evaluated.

## I. Introduction

The continuing trend toward the development of high-power density wide bandgap (WBG) power modules capable of operating simultaneously at high blocking voltages, switching frequencies and temperatures up to 200ºC calls for dynamic evolution of the insulation system of these devices. While the maximum blocking voltage for Si devices is 6.5 kV for IGBT, blocking voltages as high as 15 kV for 30 A and 24 kV for 80 A were reported for SiC IGBT devices where the volume of the 15 kV SiC IGBT is one-third that for the 6.5 kV Si-IGBT [1]. This results in the enhancement of the electric stress and increasing unacceptable PD levels that lead to the deterioration and aging of the module insulation materials and eventually failure of the module packaging [2-4].

To date, various electric field mitigation techniques ranging from geometrical techniques [5-7] to implementing functional materials as a coating layer to high-stressed regions of the package or filler added to silicone gel [8, 9] have been proposed. However, none of the introduced techniques have been fully addressed the enhanced electric stress issue within the high-density WBG power modules. In this regard, for the first time, the authors reported a highly effective *E* reduction solution through combining geometrical techniques and using dielectrics with nonlinear FDC characteristics as a coating layer to the high-stressed zones of the power module [10-16].

A power module consists of two main dielectrics, one to provide electrical insulation between the HV counterparts and the heat sink, another to protect the module components against environmental impacts as well as to prevent PDs in air. For the former one, a ceramic dielectric such as Aluminum Nitride (AlN) considered in this paper or Aluminum Oxide ($Al_2O_3$) is preferred, while for the latter one, commonly a soft polymeric encapsulation dielectric such as silicone gel is chosen.

The superior electrical and thermal characteristics of WBG semiconductors allow the power devices made from these materials to operate under voltages pulses with frequencies up to a few MHz. Furthermore, the outstanding features of WBG devices will not be diminished at temperatures up to a few hundred degrees. Accordingly, the insulation materials of the power module would experience a harsh environment under the aforementioned conditions. Hence, like all dielectrics, the dielectric characteristics–relative permittivity ($\varepsilon_r$) and ac electric conductivity ($\sigma_{ac}$)–of silicone gel and AlN may be frequency and temperature-dependent. In our previous studies [10-16], the influence of temperature and frequency on the mentioned dielectric characteristics were ignored. In [17], the authors studied the influence of temperature and frequency in 22ºC-160ºC and 50 Hz-1 kHz on the proposed electric field mitigation solution. Using the measuring data provided in the literature regarding temperature- and frequency- dependency of relative permittivity and ac electrical conductivity of silicone gel, AlN, and nonlinear FDC dielectrics, we showed that the proposed field grading technique was highly efficient within the studied range, and the electric field at the high-stressed regions of the module, where nonlinear FDC coating was active remained invariant. In this work, we extend our studies up to 250ºC and 1 MHz.

## II. Frequency, Temperature Dependency of $\varepsilon_r$ and $\sigma_{ac}$

### A. AlN Substrate

The dielectric characterization of ceramic substrates for power modules such as AlN and $Al_2O_3$ at high temperatures and high frequencies have been reported in very few studies.

In [18], $\varepsilon_r$ and $\sigma_{ac}$ are determined through measuring the complex impedance of a metalized ceramic substrate with a thickness of 1 mm under an ac voltage of 10 V and over a frequency range of 100 mHz-10 MHz from the room temperature to 450ºC, shown in Figs. 1 and 2.

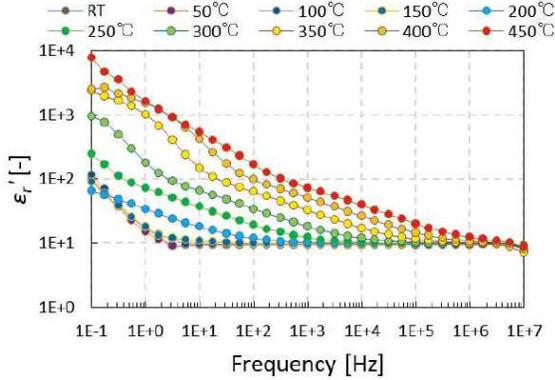

Fig. 1. $\varepsilon_r$ of AlN for different frequencies and temperatures [18].

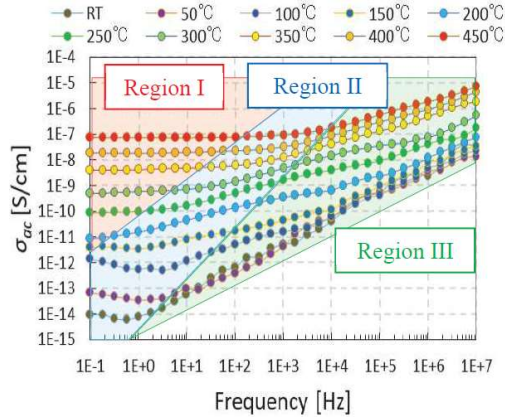

Fig. 2. $\sigma_{ac}$ of AlN for different frequencies and temperatures [18].

To evaluate the proposed field mitigation technique, simulations were performed for a base case ([22°C, 50 Hz]) and four other case studies (Case 1: [160°C, 1 kHz], Case 2: [250°C, 10 kHz], Case 3: [250°C, 100 kHz] and Case 4: [250°C, 1 MHz]).

*B. Silicone gel*

The dielectric properties of silicone gel which is widely used as an encapsulating material for power modules have been studied in [19, 20]. In this paper, simulations for the base case and Case 1 have been performed based on the experimental data from [19], and for Cases 2-4, the values of $\varepsilon_r$ and $\sigma_{ac}$ have been adopted from [20].

In [19], the measurement of $\varepsilon_r$ and $\sigma_{ac}$ over a wide frequency range up to 1 kHz and for several temperatures up to 160ºC was performed. In [20], however, $\varepsilon_r$ and $\sigma_{ac}$ for only one temperature, 250ºC, and over a frequency range of 0.1 Hz to 1 MHz was reported.

The measurement data of $\varepsilon_r$ ($\varepsilon'$) and $\varepsilon''$ (dielectric loss) reported in [20] are shown in Figs. 3 and 4, respectively. $\sigma_{ac}$ of the sample is related to the dielectric loss by

$$\sigma_{ac} = \omega \varepsilon_0 \varepsilon'' \qquad (1)$$

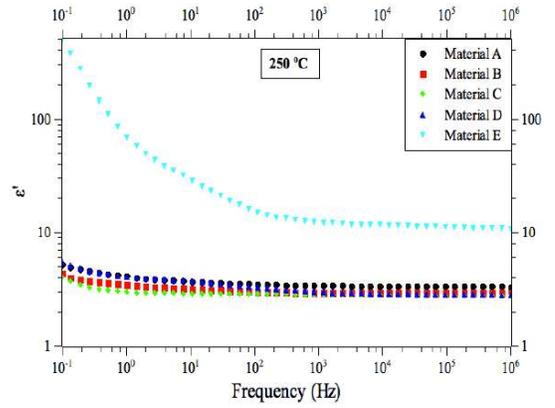

Fig. 3. $\varepsilon_r$ for silicone gel versus frequency at 250°C [20].

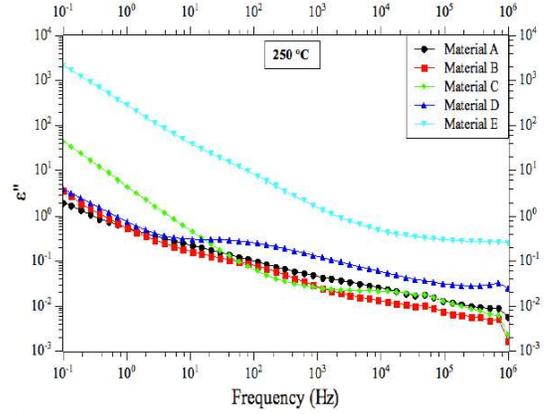

Fig. 4. $\varepsilon''$ for silicone gel versus frequency at 250°C [20].

The measurement data from Figs. 1-4 for the case studies are presented in Table I.

TABLE I. PARAMETERS USED FOR SIMULATIONS

| Case | AlN $\varepsilon_r$ | AlN $\sigma_{ac}$ (S/m) | Silicone gel $\varepsilon_r$ | Silicone gel $\sigma_{ac}$ (S/m) |
|---|---|---|---|---|
| Base case | 9.43 | $5 \times 10^{-11}$ | 2.72 | $1.1 \times 10^{-14}$ |
| Case 1 | 10.02 | $1 \times 10^{-8}$ | 2.05 | $7.4 \times 10^{-11}$ |
| Case 2 | 10.01 | $3.9 \times 10^{-7}$ | 2.88 | $1.11 \times 10^{-8}$ |
| Case 3 | 9.8 | $8.9 \times 10^{-7}$ | 2.82 | $7.34 \times 10^{-8}$ |
| Case 4 | 9.6 | $4 \times 10^{-6}$ | 2.87 | $1.31 \times 10^{-7}$ |

*C. Nonlinear FDC Material*

Nonlinear FDC composites made from the inclusion of semiconductive or blend of oxide fillers such as SiC, $TiO_2$, $SiO_2$ with an insulation matrix have been widely used for the protection of cable accessories and motor stator bars against corona in air. The new FDC composites based on microvaristor fillers such as ZnO microvaristor developed recently has received a lot of interest because their nonlinearity characteristics can be engineered very well for different applications. For the first time, the field grading feature of FDC materials with ZnO microvaristors was studied in [8]. The measuring data regarding the influence of temperature on the resistivity of a ZnO/silicone (35 vol %) compound at 22°C and 90°C demonstrate a temperature-independent resistivity for the composite [21]. It is due to the neutralizing effect of thermal expansion of the polymer (silicon) and simultaneous thermally-induced filler transportation within the polymer matrix. The field dependency of the conductivity of the ZnO/silicon compound regarding the measuring data in [21] can be fitted by

$$\sigma(E) = ae^{bE} \quad (2)$$

where $(a, b)$ for 22°C and 90°C are (4.83×10$^{-12}$, 8.552×10$^{-6}$), and (3.092×10$^{-11}$, 5.272×10$^{-6}$), respectively. As discussed, due to the rather temperature-independent resistivity of the compound, the conductivity is considered the same as it for 90°C for Cases 1-4. However, due to the lack of information on the temperature-dependency of $\varepsilon_r$ and frequency-dependency of the microvaristor compounds conductivity, $\varepsilon_r$ and $\sigma(E)$ are considered to be independent of temperature and frequency, respectively. The reproducing data from [21] for the base case and four other cases are summarized in Table II.

TABLE II. $\sigma(E)$ PARAMETERS AND $\varepsilon_r$ FOR DIFFERENT CASES [21].

| Case | a | b | $\varepsilon_r$ |
|---|---|---|---|
| Base case | $4.83 \times 10^{-12}$ | $8.552 \times 10^{-6}$ | 11.47 |
| Case 1 | $3.092 \times 10^{-11}$ | $5.272 \times 10^{-6}$ | 11.11 |
| Case 2 | $3.092 \times 10^{-11}$ | $5.272 \times 10^{-6}$ | 10.98 |
| Case 3 | $3.092 \times 10^{-11}$ | $5.272 \times 10^{-6}$ | 10.9 |
| Case 4 | $3.092 \times 10^{-11}$ | $5.272 \times 10^{-6}$ | 10.86 |

## III. MODELING AND SIMULATION RESULTS

Electric stress is higher in air bubbles formed in silicone gel or defects within the substrate, as well as around sharp metallization edges and protrusion beneath the metallization layer formed by active metal brazing (AMB) of the metal to the ceramic, and at triple points where metallization layer, silicone gel, and ceramic meet.

As a geometrical technique to reduce the locally enhanced electric stress at triple points, the protruding substrate shown in Fig. 5 was introduced [7].

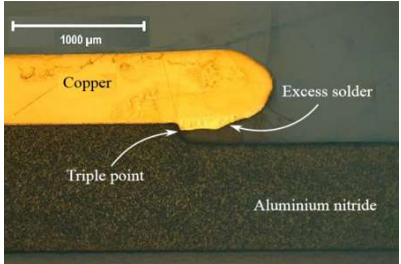

Fig. 5. Cross-section of the protruding substrate [7].

Although the protruding substrate could significantly mitigate the high electric stress within the ceramic substrate, it has a negligible influence on silicone gel. Thus, to mitigate the high electric stress in silicone gel, the high-stressed region of the module–around HV electrode edges and AlN rims–is coated by a nonlinear FDC layer as schematically shown in Fig. 6.

Dealing with FEM simulations, note that the accuracy of results is profoundly affected by the defined mesh resolution, in particular, in high-stressed regions. To this end, three different meshing zones were considered. As shown in Fig. 7a, all sharp edges are incorporated in zone 1 with extremely mesh density with the maximum element size of 4 μm. Zone 2 and zone 3 with a maximum element size of 10 μm and 400 μm for meshing, respectively, are defined for the remaining regions of the module. Moreover, $E$ values are considered along measuring lines L1 (in AlN), L2 (in silicone gel), and L3 (in FDC layer) located at a distance of 15 μm from interfaces as shown in Fig. 7b.

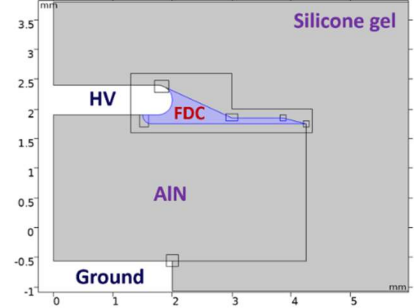

Fig. 6. The geometry considered for simulations.

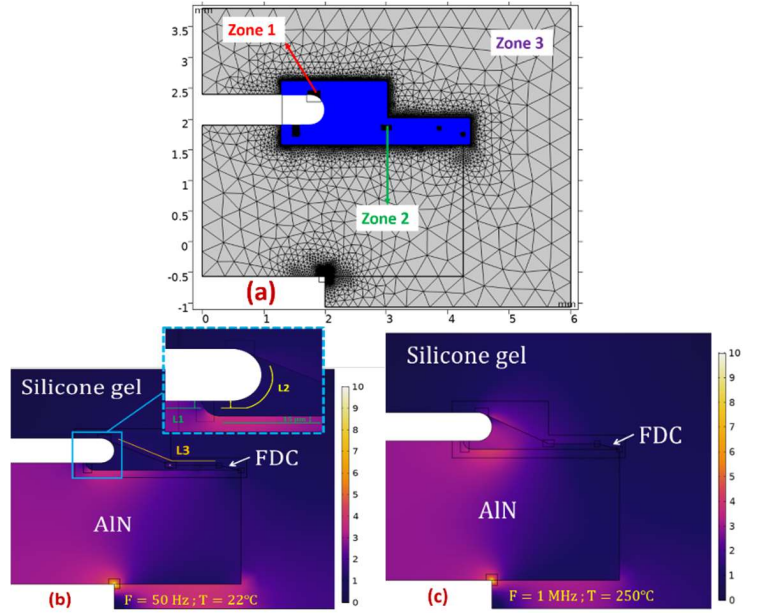

Fig. 7. (a) Meshing strategy, (b) Measuring lines and electric field distribution for the base case, (c) Electric field distribution for Case 4.

For a blocking voltage of 6.5 kV and to reproduce the PD measurement step condition from the IEC 61287-1, a 50/60 Hz ac voltage with a magnitude of $1.1U_b$=7.15 kV ($U_b$: blocking voltage) is applied to top metallization while the bottom metallization is grounded. Table III shows the maximum $E$ ($E_{max}$) obtained from simulations for the case studies.

TABLE III. $E_{max}$ ALONG L1-L3 FOR CASE STUDIES.

| Case | L1 (kV/mm) | L2 (kV/mm) | L3 (kV/mm) |
|---|---|---|---|
| Base case | 3.91 | 1.17 | 3.49 |
| Case 1 | 3.90 | 1.98 | 2.34 |
| Case 2 | 4.10 | 2.39 | 2.66 |
| Case 3 | 4.24 | 2.74 | 2.93 |
| Case 4 | 4.31 | 3.11 | 3.16 |
| Case 5 | 8.48 | 10.54 | 4.82 |

Seen from Table III, for 250ºC (Cases 2-4) $E_{max}$ along lines L1-L3 increases (by 5% in AlN, 30% in nonlinear FDC later, and 19% in silicone gel) by increasing the frequency from 10 kHz to 1 MHz. Case 5 shown in Table III is the same as Case 4 (250 ºC, 1 MHz) however without applying nonlinear FDC layer. Compared to Case 4, nonlinear FDC layer still works well at a high temperature of 250ºC under 1 MHz, leading to an $E$ reduction of 49% in AlN and 71% in silicone gel. Comparing the base case (22°C, 50 Hz) with Case 4 (250°C, 1 MHz) shows that $E$ values increase by 10% in AlN and 165% in nonlinear FDC layer, and decreases by 9% in silicone gel. From these results and regarding the availability of measurement data for $\varepsilon_r$ and $\sigma_{ac}$ for silicone gel [19, 20], AlN [18], and other ceramic substrate materials reported in the literature, it is recommended to use the values of the mentioned parameters at working conditions, especially for high temperatures and frequencies to obtain more accurate and reliable values of $E$. This may especially be important when comparing different $E$ reduction techniques. Note that PDs in air-filled voids/bubbles within silicone gel and AlN or in air-filled spaces formed at silicone gel/AlN interface were not modeled in this paper, where high temperatures and especially high frequencies will accelerate PD activities and in turn insulation degradation as shown in [22-26]. Therefore, comparing cases in Table III should be viewed on the only $E$ side and should not be interpreted on the insulation degradation side.

## IV. CONCLUSION

In this paper, it was shown that $E$ values obtained from simulations assuming the permittivity and ac electrical conductivity of the dielectrics used in power electronics modules at room temperature and 50/60 Hz ac sinusoidal voltage may not valid for high temperatures and frequencies. Models and simulations were developed and done in COMSOL Multiphysics. Using the FEM model for an efficient E reduction method via combining protruding substrate with a nonlinear FDC layer applied to high electric stress areas, the influence of temperature and frequency on $E$ values were studied. Using measuring data, studies were done for a temperature range of 22-250°C and a frequency range of 50 Hz-1 MHz. The results indicated that the mentioned E reduction method works well, even at 250°C and 1 MHz.